\UseRawInputEncoding
\documentclass[journal]{IEEEtran}
\usepackage{cite}
\usepackage{ifpdf}
\usepackage{amsmath}
\usepackage{amsfonts}
\usepackage{amssymb}
\usepackage{array}
\usepackage{graphicx,bm}
\usepackage{graphicx}
\usepackage{enumitem}
\usepackage{cleveref}
\usepackage[normalem]{ulem}
\usepackage{tikz}
\usetikzlibrary{automata, positioning, arrows,shapes.multipart,fit}
\usetikzlibrary{decorations.shapes}
\usepackage[labelfont=bf]{caption}
\usepackage{multirow}
\usepackage{graphicx,graphics,subcaption,color}

\newcommand{\norm}[1]{||#1||_1}
\newcommand{\normtwo}[1]{||#1||_2}

\renewcommand{\b}{\mathbf{b}}
\newcommand{\m}{\mathbf{m}}
\renewcommand{\j}{\mathbf{j}}

\newcommand{\thet}{\bm{\theta}}

\newcommand{\y}{\mathbf{y}}
\renewcommand{\H }{\mathbf{H}}
\newcommand{\z}{\mathbf{z}}

\newcommand{\W}{\mathbf{W}}

\newcommand{\0}{\mathbf{0}}

\newcommand{\x}{\mathbf{x}}

\renewcommand{\t}{\mathbf{t}}

\renewcommand{\v}{\mathbf{v}}

\renewcommand{\b}{\mathbf{b}}

\renewcommand{\i}{\mathbf{i}}

\renewcommand{\vec}[1]{{\rm{vec}}\(#1\)}

\renewcommand{\vec}[1]{\mathbf{#1}}

\newcommand{\RNum}[1]{\uppercase\expandafter{\romannumeral #1\relax}}

\begin{document}

\title{Green DetNet: Computation and Memory efficient DetNet using Smart Compression and Training}
\author{ Nancy Nayak$^{1}$,\hspace{4pt} Thulasi Tholeti$^{1}$, \hspace{4pt} Muralikrishnan Srinivasan$^{2}$, \hspace{4pt} Sheetal Kalyani$^{1}$\\
    \thanks{1. Nancy Nayak, Thulasi Tholeti and Sheetal Kalyani are with the Dept. of Electrical Engineering, Indian Institute of Technology, Madras, India.
			(Emails:{ee17d408@smail,ee15d410@ee,skalyani@ee}.iitm.ac.in).}
		\thanks{2. Muralikrishnan Srinivasan is with ETIS UMR8051, CY Université, ENSEA, CNRS, Cergy, France.
			(Email:muralikrishnan.srinivasan@ensea.fr)}
} 
\maketitle
\begin{abstract}
    This paper introduces an incremental training framework for compressing popular Deep Neural Network (DNN) based unfolded multiple-input-multiple-output (MIMO) detection algorithms like DetNet. The idea of incremental training is explored to select the optimal depth while training. To reduce the computation requirements or the number of FLoating point OPerations (FLOPs) and enforce sparsity in weights, the concept of structured regularization is explored using group LASSO and sparse group LASSO. Our methods lead to an astounding $98.9\%$ reduction in memory requirement and $81.63\%$ reduction in FLOPs when compared with DetNet without compromising on BER performance.
\end{abstract}


\section{Introduction}

\par In recent times, deep learning has succeeded tremendously in solving several complex data-driven problems. In particular, deep learning approaches are widely used in wireless communications to tackle problems such as signal detection \cite{farsad2017detection,dorner2018deep}, sparse signal recovery \cite{gregor2010learning}, end-to-end communication \cite{oshea2017deep,raj2018backpropogating,raj2020design}, etc. There has been a significant focus on developing deep neural networks (DNN) by unfolding existing iterative algorithms \cite{hershey2014deep,he2020envelope,monga2021algorithm}. One such algorithm is the DetNet proposed in \cite{samuel2019learning} to address the problem of detection in multiple-input-multiple-output(MIMO) systems as an alternative to classical detection techniques like maximum likelihood (ML)  \cite{he2020model,samuel2019learning,wei2020learned}.  
\par When an unfolded algorithm is implemented, each iteration is represented by a network layer and hence requires additional memory and computation. Specifically, in DetNet, each iteration results in the storage of approximately $26000$ parameters for a $30 \times 20$ channel matrix. Although DNN-based methods typically employ large-scale memory and computations, stringent memory and computation constraints arise while deploying such detection algorithms in small devices, especially low-power mobile devices and the Internet of Things-devices (IoT-devices) in 6G network. Besides, building a sustainable green network by achieving energy efficiency while supporting ultra-high data rate and ultra-low latency is one of the technical objectives of 6G networks \cite{huang2019survey}. Therefore, to implement deep learning algorithms in the edge devices with minimal memory and computation, compressing the networks by reducing the number of parameters/layers is of utmost importance \cite{park2018nncompactor,hu2021iterative}.
\par Pruning and quantization of weights are time-tested methods to reduce the memory requirements \cite{han2015deep}. However, such compression methods may lead to a decrease in the performance (as demonstrated in Section \ref{sec:numres}). Smartly pruning the weights of networks instead of simply thresholding them without pre-processing would lead to compression without significant loss in performance. A popular method of pruning is to sparsify the network's weights using regularization   \cite{han2015learningb,alvarez2016learning}. Though pruning weights reduce the memory requirements, unless pruning is done in a structured fashion, the computation requirements (number of FLoating point OPerations (FLOPs) utilized) remain the same \cite{anwar2015fixed}. Therefore, works such as \cite{wen2016learning,gale2019state} used structured sparsity, to reduce the network complexity. More recently, the authors of \cite{mohammad2020complexity} proposed WeSNet, where the weights are multiplied by a monotonic weight scaling profile function. Here, the network is pruned based on the magnitude of the weights after scaling. Furthermore, the authors have also proposed a sparsity-inducing regularization to reduce the complexity of DetNet.
\par In addition to pruning the parameters of the network with a certain depth, it is also of immense interest to determine the optimal number of layers. The number of layers that one should use has always been a subjective choice in most applications, where the user chooses the number of layers by trial and error. Therefore, in this work, we develop the following intelligent strategies to obtain a model with both minimum depth and minimum complexity without compromising on the required performance:
\begin{enumerate}
    \item \textbf{Incremental training:} Here, the weights of a small portion of the network are trained before adding the next set of layers to ensure minimum depth without a decrease in performance. In our results, we show that using \textit{incremental training enables our network to operate with just 40 layers. This saves 50 layers worth of complexity at the training stage itself.} Then further reductions are made using the regularization and pruning methods.
    \item \textbf{Regularization to induce sparsity:} We propose to apply a combination of structured and unstructured sparsity to achieve maximum benefits in terms of complexity reduction. We show that the proposed method uses $\sim 1\%$ of the memory and $<20\%$ of the FLOPs required by DetNet with almost no degradation in performance.
\end{enumerate}


\section{DetNet: Existing approaches} 
We first revisit the DetNet architecture proposed in \cite{NeevsamueldeepMIMO} and then study its memory and computation requirements.
\label{sec:detnet}

\subsection{System Model}
In a MIMO detection system model, the received vector $\vec y \in \mathbb{R}^{N}$ is given by
\begin{equation}\label{linearmodel1}
\vec y = \vec H \vec x + \vec n,
\end{equation}
where $\vec H \in \mathbb{R}^{N \times K}$ is the channel matrix, $\x$ is a vector from a constellation $\mathbb{S}$ and
$\vec n$ is the additive white Gaussian noise (AWGN) with variance $\sigma^2$. $N$ refers to the number of receivers and $K$ refers to the number of transmitters. The $k$th layer of DetNet has the following architecture \cite{NeevsamueldeepMIMO}:
\begin{eqnarray}\label{DetNetArchitecture}
 \z_{k} &=& \rho\left(\W_{1k} \left[
 \begin{array}{c}
   \H^{T}\y \\   \hat\x_{k} \\ \H^{T} \H \hat \x_{k} \\ \v_{k} 
 \end{array} \right] +\b_{1k}\right), \quad k=1,..,L \nonumber\\
 \hat\x_{k+1}&=&\psi_{t_k}\left(\W_{2k}\z_k+\b_{2k}\right),\nonumber\\
 \v_{k+1}&=&\W_{3k}\z_k+\b_{3k}, \nonumber\\
 \hat\x_1 &=& \0.
\end{eqnarray}
Here, $\rho$ is the rectified linear unit and $\psi_t(\cdot)$ is a piece-wise linear soft sign operator.
The parameters that are optimized during the learning phase are:
\begin{eqnarray}
 \thet=\left\{\W_{1k},\b_{1k},\W_{2k},\b_{2k},\W_{3k},\b_{1k},\t_{k}\right\}_{k=1}^L.
\end{eqnarray}
The loss to be minimized is defined as 
\begin{eqnarray} 
 l(\x; \hat \x_{\thet} (\H,\y)) = \sum_{k=1}^L log(k) \dfrac{\norm{\x - \hat\x_k}^2}{\norm{\x - \Tilde{\x}}^2},
\end{eqnarray}
where $\Tilde{\x} = (\H^T \H )^{-1} \H^T \y$ is the standard decoder, $\lambda$ is the regularization parameter. The final estimate is defined as $\hat\x_{\thet}\left(\y,\H\right)=sign(\hat \x_L)$.

\subsection{Related works}
DetNet has gained immense popularity recently, and there has been an interest in reducing the complexity of DetNet.

The authors of \cite{mohammad2020complexity} proposed a method (named WeSNet-$x\%$) where the weights are multiplied by a monotonic weight scaling profile function and $x\%$ of the network is pruned based on the magnitude of the weights after scaling. As the profile function is monotonic, this process results in structured sparsity, saving memory, and computation. WeSNet pruned the layer weights with a lesser value of profile function in each layer while leaving the part of the layer weights with high profile function untouched. Therefore a part of the input can be ignored entirely for computation. Also, WesNet can prune up to $40\%$ of the weights without any degradation in the performance compared to DetNet.
\par In contrast, we train the network to operate with a smaller number of layers by exploiting the incremental training approach. The network with reduced layers is then trained with a regularized loss function to prune the weights further. As our method allows the pruning of neurons without removing the entire layer, the network has more degrees of freedom to learn the essential parameters. We compare the sparsity and computation reduction with WesNet in Section \ref{sec:numres}.

\section{Proposed complexity reduction}
\label{sec:detnetpropchanges}
To choose the minimum depth and to prune the weights of the network, we propose the following three methods:

\subsection{Incremental training}
In the first approach, training is done in steps of $T_{step}$ layers. We term the training of $T_{step}$ layers as a single step of incremental training. At the end of $t$th step of the incremental training, the first $tT_{step}$ layers are frozen and not updated anymore. Therefore, during the $(t+1)$th step, only the layers $k= tT_{step}+1,.., (t+1)T_{step}$ are trained. That is, 
\begin{align}
 \thet&=\{\W_{1k},\b_{1k},\W_{2k},\b_{2k},\W_{3k},\b_{1k},\t_{k}, \nonumber \\
 & \hspace{2cm} \forall k= tT_{step}+1,..., (t+1)T_{step} \}
\end{align}
are the trainable parameters. Training is halted once the reduction in loss with additional layers is negligible. There are multiple advantages to incremental training:
\begin{enumerate}
    \item \textit{The number of layers is no longer a hyperparameter that has to be tuned.} It can be rounded to the nearest value of $T_{step}$ during incremental training. This also addresses the model/depth selection problem while training. Now one can choose to stop adding layers once a required BER is achieved. 
    \item \textit{Significant memory and computation savings are obtained corresponding to the layers that were not implemented.} We know that each layer of DetNet corresponds to 26,000 parameters. Therefore, if incremental training results in 50 layers instead of the original 90 layers, memory and computation involved for $1.04 \times 10^6$ parameters are saved.
    \item \textit{Vanishing/exploding gradient problems can be avoided.} While training deep networks, whenever the gradient's magnitude diminishes/increases uncontrollably, the problem of vanishing or exploding gradient is encountered. As incremental training only deals with training a small portion of the network, the issue is avoided. 
\end{enumerate}
Therefore, incremental training picks the network of a lower depth during the training process itself. Once the depth is chosen, the following methods can be used to obtain the least complex network given a certain depth.

\subsection{Regularized training for inducing sparsity}
Pruning weights with comparatively lower magnitudes is a popular idea in machine learning. Hence, we propose to regularize the weights so that more entries have lower magnitudes and can be pruned. Therefore, a cost on the magnitude of the weights is imposed while training. The loss to be minimized can be hence defined by 
\begin{align}
 l(\x; \hat \x_{\thet} (\H,\y)) &= \sum_{k=1}^L log(k) \dfrac{\norm{\x - \hat\x_k}^2}{\norm{\x - \Tilde{\x}}^2} + \nonumber \\
 & \hspace{-1cm} \lambda \sum_{k=1}^L (\norm{\W_{1k}}+\norm{\W_{2k}} + \norm{\W_{3k}} ). \label{eqn:detnetLoss}
\end{align}
Note that the term  $\sum_{k=1}^L \left(\norm{\W_{1k}}+\norm{\W_{2k}}+\norm{\W_{3k}} \right)$ is an additional regularization term that acts as a penalty to regulate the magnitude of the weights of the network and is absent in the popular DetNet architecture. Note that $\lambda=0$ denotes the vanilla DetNet architecture, while $\lambda >0$ is henceforth called R-DetNet. Here, the regularization is unstructured, as the entries in the weight matrix that are regularized can be random. When pruned, they are replaced with zero, thereby saving the memory for storing the parameter. However, as the zero entries' locations are not structured, this regularization does not save computation. To further reduce the number of FLOPs, one requires structured regularization.

\ifCLASSOPTIONonecolumn

\tikzset{%
  every neuron/.style={
    circle,
    draw,
    minimum size=0.8cm
  },
  neuron missing/.style={
    draw=none, 
    scale=4,
    text height=0.333cm,
    execute at begin node=\color{black}$\vdots$
  }
}
\tikzset{decorate sep/.style 2 args=
{decorate,decoration={shape backgrounds,shape=circle,shape size=#1,shape sep=#2}}}

\begin{figure}[ht]
\centering
\begin{tikzpicture}[scale = 0.7,x=2cm, y=2cm, >=stealth]

\foreach \m/\l [count=\y] in {1,2,missing,3}
  \node [every neuron/.try, neuron \m/.try] (input-\m) at (0,2.5-\y) {};

\foreach \m [count=\y] in {1,2,missing,3}
  \node [every neuron/.try, neuron \m/.try ] (hidden1-\m) at (2,2.5-\y) {};

\foreach \m [count=\y] in {1,2,missing,3}
  \node [every neuron/.try, neuron \m/.try ] (hidden2-\m) at (4,2.5-\y) {};

\foreach \l [count=\i] in {1,2,n}
  \draw [<-] (input-\i) -- ++(-1,0)
    node [above, midway] {$x_\l$};

\draw[rotate around={45:(0.35,1.30)},red,thick,dashed] (0.35,1.30) ellipse (0.75cm and 0.25cm);
\draw[rotate around={45:(2.35,1.30)},red,thick,dashed] (2.35,1.30) ellipse (0.75cm and 0.25cm);

\draw[rotate around={135:(0.35,-1.30)},red,thick,dashed] (0.35,-1.30) ellipse (0.75cm and 0.25cm);
\draw[rotate around={135:(2.35,-1.30)},red,thick,dashed] (2.35,-1.30) ellipse (0.75cm and 0.25cm);

\draw[rotate around={90:(0.38,0.4)},red,thick,dashed] (0.38,0.4) ellipse (0.75cm and 0.25cm);
\draw[rotate around={90:(2.38,0.4)},red,thick,dashed] (2.38,0.4) ellipse (0.75cm and 0.25cm);

\foreach \i in {1,...,3}
  \foreach \j in {1,...,3}
    \draw [->] (input-\i) -- (hidden1-\j);
    
\foreach \i in {1,...,3}
  \foreach \j in {1,...,3}
    \draw [->] (hidden1-\i) -- (hidden2-\j);

\draw[decorate sep={1.25mm}{4mm},fill] (4.5,1.5) -- ++(0.8,0);

\draw[decorate sep={1.25mm}{4mm},fill] (4.5,0.5) -- ++(0.8,0);

\draw[decorate sep={1.25mm}{4mm},fill] (4.5,-1.5) -- ++(0.8,0);


\end{tikzpicture}
   \caption{Grouping of weights for structured regularization}
    \label{fig:grouping}
\end{figure}
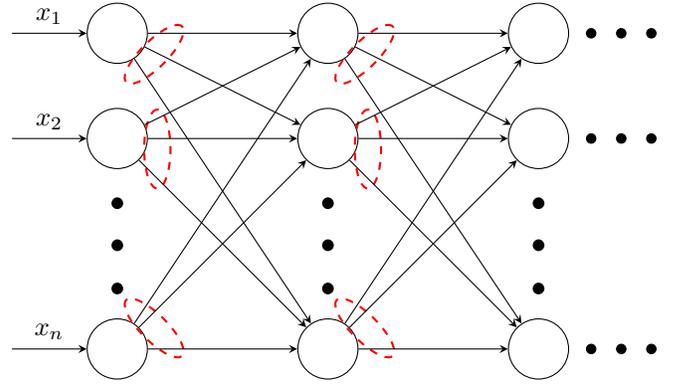
\fi
Group LASSO (GL) technique was proposed in \cite{meier2008group} in the context of logistic regression in which weights are grouped, and an entire group of weights is regularized as a unit. It has been successfully used for pruning feed-forward networks by the authors of \cite{groupLASSO} for various regression and classification tasks. Whenever a group is regularized by forcing the group entries to zero, a reduction in FLOPs is achieved. To do so, consider $\Tilde{\W}_{ik} = [\W_{ik} \quad \b_{ik}]$, where $\Tilde{\W}_{1k}^{i}$ denotes the $i^{th}$ column of the matrix $\Tilde{\W}_{1k}$. We propose using each column as a group (highlighted by weights encircled by red dots in Fig. \ref{fig:grouping}) while performing the group LASSO regularization. Although group LASSO saves both memory and computation by setting all the weights in a group to zero, there may be individual weight entries in other non-zero groups that could be further pruned resulting in increased memory savings in addition to the computation saving obtained from Group LASSO. In statistical literature, \cite{simon2013sparse} proposed Sparse Group LASSO (SGL), which tries to achieve both element wise and group sparsity. We propose to use the SGL penalty function as a regularizer for DetNet to reduce both memory and FLOPs efficiently. The loss function to be minimized for this method is: 
\begin{align}
 l(\x; \hat \x_{\thet} (\H,\y)) &= \sum_{k=1}^L log(k) \dfrac{\norm{\x - \hat\x_k}^2}{\norm{\x - \Tilde{\x}}^2} + \nonumber \\
 & \hspace{-1.2cm} \lambda_1 \sum_{k=1}^L \left(\sum_{i=1}^{I_1}\normtwo{\Tilde{\W}_{1k}^{i}}+\sum_{i=1}^{I_2}\normtwo{\Tilde{\W}_{2k}^{i}}+\sum_{i=1}^{I_3}\normtwo{\Tilde{\W}_{3k}^{i}} \right) \nonumber \\
 & \hspace{-1.2cm} + \lambda_2 \sum_{k=1}^L (\norm{\W_{1k}}+\norm{\W_{2k}} + \norm{\W_{3k}} ). \label{eqn:detnetLossSGL}
\end{align}
The two regularization parameters $\lambda_1$ and $\lambda_2$ should be tuned independently. This method provides higher savings on memory than simply using group LASSO while imposing structural sparsity. Incremental training can be used with sparsity inducing regularization to obtain a computation and memory-efficient variant of DetNet. 

\ifCLASSOPTIONtwocolumn

\fi

\section{Numerical Results} \label{sec:numres}
In all the experiments, we have considered a MIMO channel with input size $K=20$ and output size $N=30$ similar to \cite{samuel2019learning}. The other simulation settings are similar to the experimental setup of DetNet proposed in \cite{NeevsamueldeepMIMO}.
The input vector $\x$ is drawn from a BPSK constellation, and $20000$ mini-batches of size $1000$ are generated for each SNR value; we consider SNRs from 0-15dB. The entries of the channel matrix $\H$ are sampled from $\mathcal{N}\left(0,1\right)$. The network is trained using Adam Optimizer \cite{kingma2014adam}; the learning rate decays exponentially starting with an initial learning rate of $10^{-4}$, a decay factor $0.97$ and decay step-size of $1000$. Each floating-point addition/multiplication is considered as one FLOP. The memory requirement of a network is calculated considering 32 bits for all floating-point parameters of the network that need to be saved. 
\par For a particular layer $k$, the weights lesser than a threshold $W_{thr}^k$ are pruned. Using $W_{thr}^k= \eta \times W_{max}^k,$
where $W_{max}^k$ is the maximum absolute value of the weights in layer $k$ and $\eta$ is a user-defined parameter that defines the extent of pruning. A higher $\eta$ will result in a higher threshold, and hence more parameters are being pruned. Similarly, for group LASSO, groups below a certain threshold are pruned. In the case of the proposed sparse group LASSO for incremental training, we employ two fractions: $\eta_1$ to obtain the threshold for pruning the groups and $\eta_2$ for further pruning the weights.

\begin{table}[ht]
    \centering
    \begin{tabular}{|c|c|c|c|}
    \hline
    Architecture & Memory (MB) & FLOPs & BER at $12$ dB \\
    \hline
    \hline
        DetNet & $9.190$ & $4.90e6$ & $0.0012$\\
        \hline
        R-I-DetNet(SGL)-30L  & $0.0766$  & $0.56e6$ & $0.00129$\\
        \hline
        R-I-DetNet(SGL)-40L  & $0.0939$ & $0.72e6$ & $0.001234$\\
        \hline
        R-I-DetNet(SGL)-50L  & $0.1072$  & $0.89e6$ & $\mathbf{ 0.001200}$ \\
        \hline
        R-I-DetNet(SGL)-60L  & $0.1225$  & $1.05e6$ & $0.001151$\\
        \hline
        R-I-DetNet(SGL)-70L  & $0.1429$  & $1.22e6$ & $0.001279$\\
        \hline
        R-I-DetNet(SGL)-80L & $0.1634$  & $1.39e6$ & $0.001380$\\
        \hline
        R-I-DetNet(SGL)-90L  & $0.1838$  & $1.57e6$ & $0.001343$\\
        \hline
    \end{tabular}
    \caption{Memory and computation saving in incremental training with sparse group LASSO regularization.}
    \label{tab:memcompute_act}
\end{table}

In the proposed incremental training method, instead of training a DetNet with all the $90$ layers, we train a network with a smaller number of layers (we begin with 30 layers) and add more layers only if the target BER is not achieved. Structural regularization using sparse group LASSO is incorporated during the training to save memory/computation further. For the experiments in this paper, $\lambda_1$ and $\lambda_2$ are chosen to be $0.04$. As shown in Table \ref{tab:memcompute_act}, the memory requirement of every R-I-DetNet(SGL)-$x$L network with $x$ layers is better than that of the original DetNet. Also, from Table  \ref{tab:memcompute_act}, with the addition of layers, BER gradually decreases and achieves the BER of the original DetNet $(0.0012)$ at depth $x=50$. With further addition of layers, BER improves slightly at the cost of increased memory requirement and FLOP-counts. Instead of employing an ad-hoc method to choose the depth of a network, the proposed method enables the user to choose the depth while the training progresses depending on the BER requirement of the task at hand.

\ifCLASSOPTIONtwocolumn
\begin{table}
    \centering
    \begin{tabular}[scale=0.95]{|c|c|c|c|}
    \hline
    Architecture & Memory(MB) & FLOPs & BER\\
    \hline
    \hline
        DetNet & $9.19$ & $4.9e6$ & $0.00118$\\
        \hline
        \textbf{R-I-DetNet(SGL)-50L} &  $\mathbf{0.10}$  & $\mathbf{0.9e6}$ & $\mathbf{0.00120}$\\
        \hline
        Pruned DetNet ($\eta=0.05$) & $5.05$  & $4.9e6$ & $0.03925$ \\
        \hline
        \textbf{R-DetNet} ($\eta=0.05$) &  $2.30$  & $4.9e6$ & $0.00118$\\
        \hline
        WesNet &  $5.51$ & $3.1e6$  & $0.00095$ \\
        \hline
        \textbf{R-DetNet(GL)} ($\eta=0.005$) & $4.13$  & $2.6e6$ & $0.00157$\\
        \hline
        \textbf{R-DetNet(SGL)} & $1.87$  & $2.6e6$ & $0.00115$\\
        \hline
        \textbf{R-I-DetNet-40L} & $1.02$ & $2.2e6$  & $0.00123$ \\
        \hline
    \end{tabular}
    \caption{Comparison of proposed (in bold) with existing methods for SNR$=12$dB}
    \label{tab:MACCBER}
\end{table}
\fi

\ifCLASSOPTIONonecolumn
\begin{table}[ht]
    \centering
    \begin{tabular}[scale=0.95]{|c|c|c|c|}
    \hline
    Architecture & Memory(MB) & FLOPs & BER\\
    \hline
    \hline
        DetNet & $9.19$ & $4.9e6$ & $0.00118$\\
        \hline
        \textbf{R-I-DetNet(SGL)-50L} &  $\mathbf{0.10}$  & $\mathbf{0.9e6}$ & $\mathbf{0.00120}$\\
        \hline
        Pruned DetNet ($\eta=0.05$) & $5.05$  & $4.9e6$ & $0.03925$ \\
        \hline
        \textbf{R-DetNet} ($\eta=0.05$) &  $2.30$  & $4.9e6$ & $0.00118$\\
        \hline
        WesNet &  $5.51$ & $3.1e6$  & $0.00095$ \\
        \hline
        \textbf{R-DetNet(GL)} ($\eta=0.005$) & $4.13$  & $2.6e6$ & $0.00157$\\
        \hline
        \textbf{R-DetNet(SGL)} & $1.87$  & $2.6e6$ & $0.00115$\\
        \hline
        \textbf{R-I-DetNet-40L} & $1.02$ & $2.2e6$  & $0.00123$ \\
        \hline
    \end{tabular}
    \caption{Comparison of proposed (in bold) with existing methods for SNR$=12$dB}
    \label{tab:MACCBER}
\end{table}
\fi

Table \ref{tab:MACCBER} provides a comparison of our proposed methods, namely R-DetNet, R-DetNet(GL), R-DetNet(SGL), R-I-DetNet and R-I-DetNet(SGL) (our most competitive method) with pruned DetNet and WesNet. All the experiments are performed at SNR=12dB. Similar trends can be noted for the other SNR values as well. As seen in Table \ref{tab:MACCBER}, pruning DetNet without regularization does not sparsify the network effectively. In R-DetNet, about $75\%$ of the weights are removed after pruning with an $\eta$ of $0.05$ that gives a BER of $1e-3$ approximately. However, R-DetNet does not offer any savings in terms of computation. The proposed R-DetNet(GL) has an advantage over R-DetNet in reducing the number of FLOPs by imposing a structured sparsity. At lower values of $\eta$, R-DetNet(GL) achieves BER nearly similar to that of R-DetNet while using only $53\%$ of the FLOPs. However it does not offer as much memory saving as R-DetNet.  R-DetNet(SGL) results in a highly sparse network with a memory requirement of $1.87$MB, with a considerable reduction in FLOPs (using only $53\%$) while ensuring competitive performance in terms of BER. The higher sparsity of R-DetNet(SGL) compared to WesNet can be attributed to the fact that any row/column can be chosen to be pruned, unlike WesNet, where the consecutive layer weight parameters are pruned. R-I-DetNet-40L achieves a BER of $0.00123$ after pruning which is close to that of the original network.
\ifCLASSOPTIONonecolumn
\begin{figure}[ht]
    \centering
    \includegraphics[scale=1.0]{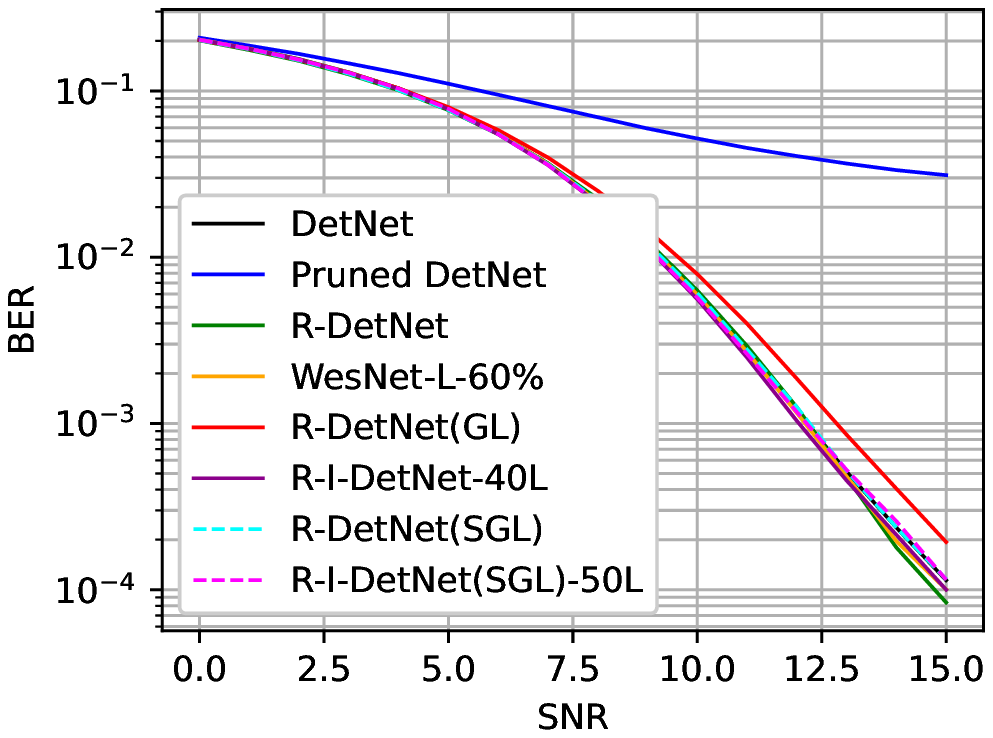} \caption{BER performance of the proposed methods compared to existing methods}
    \label{fig:BERvsSNR}
\end{figure}
\fi
Note that this performance is achieved with an astounding $88.9\%$ saving in memory and a $55.5\%$ reduction in the number of FLOPs. The savings can be further enhanced by using incremental training with sparse group LASSO i.e. our best method R-I-DetNet(SGL)-50L that requires $98.9\%$ less memory compared to DetNet; also, it requires FLOPs of $0.89e6$ compared to $4.9e6$ thus leading to $81.63\%$ lesser computation than DetNet to achieve a nearly similar BER of $\sim 0.0012$.

\ifCLASSOPTIONtwocolumn
\begin{figure}[ht]
    \centering
    \includegraphics[scale=0.8]{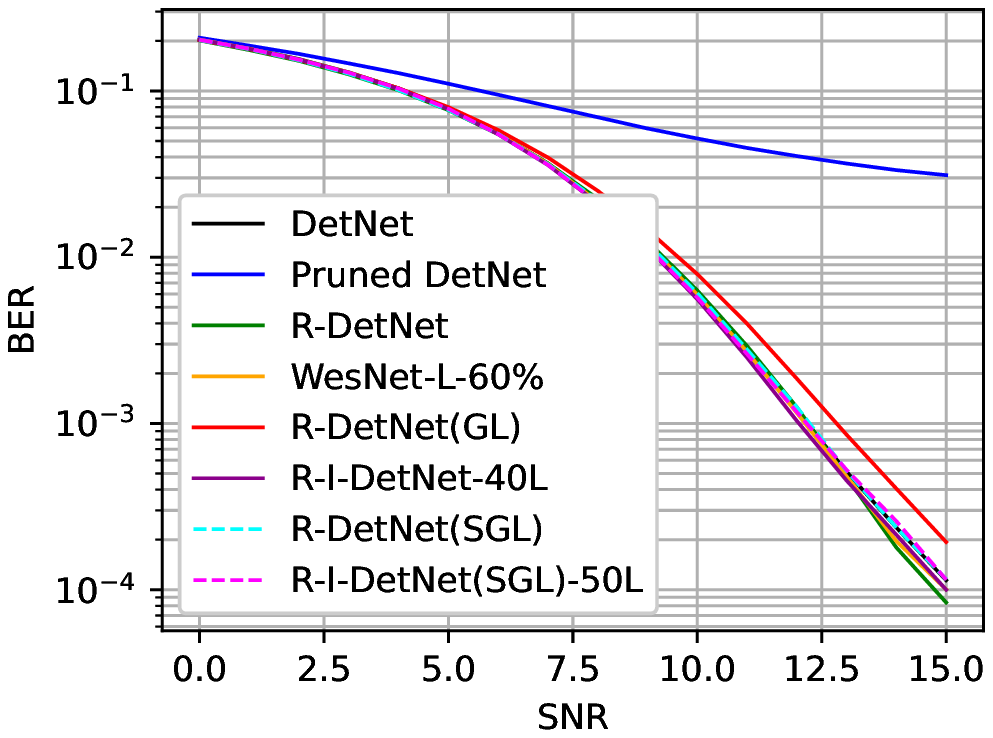} \caption{BER performance of the proposed methods compared to existing methods}
    \label{fig:BERvsSNR}
\end{figure}
\fi

Finally, we compare the BER vs SNR curves of all the methods\footnote{ To prune DetNet and R-DetNet, $\eta=0.05$ was used; for R-DetNet(GL), $\eta=0.005$ and for R-I-DetNet, $\eta=0.01$ were used.} in Fig. \ref{fig:BERvsSNR} for an SNR from $0$dB to $15$dB. For SGL, $\eta_1=0.0005$ and $\eta_2=0.01$ are used. As noted earlier, pruning DetNet without any form of regularization results in poor performance and is shown by the curve corresponding to the Pruned DetNet. At lower SNRs, it can be noted that the performance of all the other methods is similar. At higher SNRs, we observe that R-DetNet and WesNet are comparable to that of the original DetNet. Structured pruning in R-DetNet(GL) slightly increases the BER, but it is negligible compared to the gain obtained in memory and computation. Except for R-DetNet(GL), all other proposed methods including R-DetNet(SGL), R-I-DetNet-40L, and  R-I-DetNet(SGL) require significantly less memory and are computationally efficient without any degradation in BER as shown in Fig. \ref{fig:BERvsSNR}. To summarize, our method offers the possibility of a sizeable reduction in both memory and computation without any noticeable BER degradation.

\section{Conclusions} \label{sec:conc}
In this work, we focused on reducing the memory and computational requirements of DetNet during deployment utilizing different regularizing techniques in the framework of incremental training. The idea of incremental training was explored to select the optimal depth while training. It served as an effective alternative to existing ad-hoc methods of determining the necessary depth. Employing SGL regularization with incremental training resulted in a decrease in the memory requirement by $98.9\%$ and FLOPs by $81.63\%$. Hence our method is a step forward in the direction of enabling edge computation for MIMO and wireless problems. We believe that the proposed solution will lead to green/energy-efficient receivers which can be utilized for low-cost devices like IoT devices.

\section{Acknowledgements}
We would like to thank Dr. Yonina Eldar (Professor of Electrical Engineering, Weizmann institute of Science) for the valuable discussions at the onset of this work and for sharing her insights during the course of this research. 

\bibliographystyle{IEEEtran}
\bibliography{bibfile.bib}

\begin{thebibliography}{10}
\providecommand{\url}[1]{#1}
\csname url@samestyle\endcsname
\providecommand{\newblock}{\relax}
\providecommand{\bibinfo}[2]{#2}
\providecommand{\BIBentrySTDinterwordspacing}{\spaceskip=0pt\relax}
\providecommand{\BIBentryALTinterwordstretchfactor}{4}
\providecommand{\BIBentryALTinterwordspacing}{\spaceskip=\fontdimen2\font plus
\BIBentryALTinterwordstretchfactor\fontdimen3\font minus
  \fontdimen4\font\relax}
\providecommand{\BIBforeignlanguage}[2]{{%
\expandafter\ifx\csname l@#1\endcsname\relax
\typeout{** WARNING: IEEEtran.bst: No hyphenation pattern has been}%
\typeout{** loaded for the language `#1'. Using the pattern for}%
\typeout{** the default language instead.}%
\else
\language=\csname l@#1\endcsname
\fi
#2}}
\providecommand{\BIBdecl}{\relax}
\BIBdecl

\bibitem{farsad2017detection}
\BIBentryALTinterwordspacing
N.~Farsad and A.~J. Goldsmith, ``Detection algorithms for communication systems
  using deep learning,'' \emph{CoRR}, vol. abs/1705.08044, 2017. [Online].
  Available: \url{http://arxiv.org/abs/1705.08044}
\BIBentrySTDinterwordspacing

\bibitem{dorner2018deep}
S.~D{\"o}rner, S.~Cammerer, J.~Hoydis, and S.~Ten~Brink, ``Deep learning based
  communication over the air,'' \emph{IEEE Journal of Selected Topics in Signal
  Processing}, vol.~12, no.~1, pp. 132--143, 2017.

\bibitem{gregor2010learning}
\BIBentryALTinterwordspacing
K.~Gregor and Y.~LeCun, ``Learning fast approximations of sparse coding,'' in
  \emph{Proceedings of the 27th International Conference on International
  Conference on Machine Learning}, ser. ICML'10.\hskip 1em plus 0.5em minus
  0.4em\relax USA: Omnipress, 2010, pp. 399--406. [Online]. Available:
  \url{http://dl.acm.org/citation.cfm?id=3104322.3104374}
\BIBentrySTDinterwordspacing

\bibitem{oshea2017deep}
\BIBentryALTinterwordspacing
T.~J. O'Shea, T.~Erpek, and T.~C. Clancy, ``Deep learning based {MIMO}
  communications,'' \emph{CoRR}, vol. abs/1707.07980, 2017. [Online].
  Available: \url{http://arxiv.org/abs/1707.07980}
\BIBentrySTDinterwordspacing

\bibitem{raj2018backpropogating}
V.~{Raj} and S.~{Kalyani}, ``Backpropagating through the air: Deep learning at
  physical layer without channel models,'' \emph{IEEE Communications Letters},
  vol.~22, no.~11, pp. 2278--2281, Nov 2018.

\bibitem{raj2020design}
V.~Raj and S.~Kalyani, ``Design of communication systems using deep learning: A
  variational inference perspective,'' \emph{IEEE Transactions on Cognitive
  Communications and Networking}, vol.~6, no.~4, pp. 1320--1334, 2020.

\bibitem{hershey2014deep}
\BIBentryALTinterwordspacing
J.~R. Hershey, J.~L. Roux, and F.~Weninger, ``Deep unfolding: Model-based
  inspiration of novel deep architectures,'' \emph{CoRR}, vol. abs/1409.2574,
  2014. [Online]. Available: \url{http://arxiv.org/abs/1409.2574}
\BIBentrySTDinterwordspacing

\bibitem{he2020envelope}
Y.~{He}, H.~{He}, C.~K. {Wen}, and S.~{Jin}, ``Model-driven deep learning for
  massive multiuser mimo constant envelope precoding,'' \emph{IEEE Wireless
  Communications Letters}, vol.~9, no.~11, pp. 1835--1839, 2020.

\bibitem{monga2021algorithm}
V.~Monga, Y.~Li, and Y.~C. Eldar, ``Algorithm unrolling: Interpretable,
  efficient deep learning for signal and image processing,'' \emph{IEEE Signal
  Processing Magazine}, vol.~38, no.~2, pp. 18--44, 2021.

\bibitem{samuel2019learning}
N.~Samuel, T.~Diskin, and A.~Wiesel, ``Learning to detect,'' \emph{IEEE
  Transactions on Signal Processing}, vol.~67, no.~10, pp. 2554--2564, 2019.

\bibitem{he2020model}
H.~He, C.-K. Wen, S.~Jin, and G.~Y. Li, ``Model-driven deep learning for mimo
  detection,'' \emph{IEEE Transactions on Signal Processing}, vol.~68, pp.
  1702--1715, 2020.

\bibitem{wei2020learned}
Y.~Wei, M.-M. Zhao, M.~Hong, M.-J. Zhao, and M.~Lei, ``Learned conjugate
  gradient descent network for massive mimo detection,'' \emph{IEEE
  Transactions on Signal Processing}, vol.~68, pp. 6336--6349, 2020.

\bibitem{huang2019survey}
T.~Huang, W.~Yang, J.~Wu, J.~Ma, X.~Zhang, and D.~Zhang, ``A survey on green 6g
  network: Architecture and technologies,'' \emph{IEEE Access}, vol.~7, pp.
  175\,758--175\,768, 2019.

\bibitem{park2018nncompactor}
S.~{Hong}, I.~{Lee}, and Y.~{Park}, ``Nn compactor: Minimizing memory and logic
  resources for small neural networks,'' in \emph{2018 Design, Automation Test
  in Europe Conference Exhibition (DATE)}, 2018, pp. 581--584.

\bibitem{hu2021iterative}
Q.~{Hu}, Y.~{Cai}, Q.~{Shi}, K.~{Xu}, G.~{Yu}, and Z.~{Ding}, ``Iterative
  algorithm induced deep-unfolding neural networks: Precoding design for
  multiuser mimo systems,'' \emph{IEEE Transactions on Wireless
  Communications}, vol.~20, no.~2, pp. 1394--1410, 2021.

\bibitem{han2015deep}
S.~Han, H.~Mao, and W.~J. Dally, ``Deep compression: Compressing deep neural
  networks with pruning, trained quantization and huffman coding,'' \emph{4th
  International Conference on Learning Representations, {ICLR}}, 2016.

\bibitem{han2015learningb}
S.~Han, J.~Pool, J.~Tran, and W.~Dally, ``Learning both weights and connections
  for efficient neural network,'' in \emph{Advances in neural information
  processing systems}, 2015, pp. 1135--1143.

\bibitem{alvarez2016learning}
J.~M. Alvarez and M.~Salzmann, ``Learning the number of neurons in deep
  networks,'' in \emph{Advances in Neural Information Processing Systems},
  2016, pp. 2270--2278.

\bibitem{anwar2015fixed}
S.~{Anwar}, K.~{Hwang}, and W.~{Sung}, ``Fixed point optimization of deep
  convolutional neural networks for object recognition,'' in \emph{2015 IEEE
  International Conference on Acoustics, Speech and Signal Processing
  (ICASSP)}, April 2015, pp. 1131--1135.

\bibitem{wen2016learning}
W.~Wen, C.~Wu, Y.~Wang, Y.~Chen, and H.~Li, ``Learning structured sparsity in
  deep neural networks,'' in \emph{Proceedings of the 30th International
  Conference on Neural Information Processing Systems}, 2016, pp. 2082--2090.

\bibitem{gale2019state}
T.~Gale, E.~Elsen, and S.~Hooker, ``The state of sparsity in deep neural
  networks,'' \emph{CoRR}, vol. abs/1902.09574, 2019.

\bibitem{mohammad2020complexity}
A.~Mohammad, C.~Masouros, and Y.~Andreopoulos, ``Complexity-scalable
  neural-network-based mimo detection with learnable weight scaling,''
  \emph{IEEE Transactions on Communications}, vol.~68, no.~10, pp. 6101--6113,
  2020.

\bibitem{NeevsamueldeepMIMO}
N.~{Samuel}, T.~{Diskin}, and A.~{Wiesel}, ``Deep mimo detection,'' in
  \emph{2017 IEEE 18th International Workshop on Signal Processing Advances in
  Wireless Communications (SPAWC)}, July 2017, pp. 1--5.

\bibitem{meier2008group}
L.~Meier, S.~Van De~Geer, and P.~B{\"u}hlmann, ``The group lasso for logistic
  regression,'' \emph{Journal of the Royal Statistical Society: Series B
  (Statistical Methodology)}, vol.~70, no.~1, pp. 53--71, 2008.

\bibitem{groupLASSO}
J.~{Wang}, C.~{Xu}, X.~{Yang}, and J.~M. {Zurada}, ``A novel pruning algorithm
  for smoothing feedforward neural networks based on group lasso method,''
  \emph{IEEE Transactions on Neural Networks and Learning Systems}, vol.~29,
  no.~5, pp. 2012--2024, 2018.

\bibitem{simon2013sparse}
N.~Simon, J.~Friedman, T.~Hastie, and R.~Tibshirani, ``A sparse-group lasso,''
  \emph{Journal of computational and graphical statistics}, vol.~22, no.~2, pp.
  231--245, 2013.

\bibitem{kingma2014adam}
D.~P. Kingma and J.~Ba, ``Adam: {A} method for stochastic optimization,'' in
  \emph{{ICLR}}, 2015.

\end{thebibliography}

\end{document}